\documentclass{elsart5p}
\usepackage{psfrag}
\usepackage{amssymb}
\usepackage{graphicx}
\begin{document}
\begin{frontmatter}

\title{Air Shower Measurements with the LOPES Radio Antenna Array}

\vspace{4mm}
\author{A.~Haungs$^{a,\star}$,} 
\author{W.D.~Apel$^{a}$,} 
\author{J.C.~Arteaga$^{b,1}$,} 
\author{T.~Asch$^{c}$,}
\author{J.~Auffenberg$^{d}$,} 
\author{F.~Badea$^{a}$,} 
\author{L.~B\"ahren$^{e}$,} 
\author{K.~Bekk$^{a}$,}
\author{M.~Bertaina$^{f}$,} 
\author{P.L.~Biermann$^{g}$,} 
\author{J.~Bl\"umer$^{a,b}$,}
\author{H.~Bozdog$^{a}$,}
\author{I.M.~Brancus$^{h}$,} 
\author{M.~Br\"uggemann$^{i}$,} 
\author{P.~Buchholz$^{i}$,}
\author{S.~Buitink$^{e}$,} 
\author{E.~Cantoni$^{f,j}$,}
\author{A.~Chiavassa$^{f}$,} 
\author{F.~Cossavella$^{b}$,}
\author{K.~Daumiller$^{a}$,} 
\author{V.~de Souza$^{b,2}$,} 
\author{F.~Di~Pierro$^{f}$,}
\author{P.~Doll$^{a}$,}
\author{R.~Engel$^{a}$,} 
\author{H.~Falcke$^{e,k}$,} 
\author{M.~Finger$^{a}$,}
\author{D.~Fuhrmann$^{d}$,}
\author{H.~Gemmeke$^{c}$,} 
\author{P.L.~Ghia$^{j}$,}
\author{R.~Glasstetter$^{d}$,}
\author{C.~Grupen$^{i}$,} 
\author{D.~Heck$^{a}$,}
\author{J.R.~H\"orandel$^{e}$,} 
\author{A.~Horneffer$^{e}$,} 
\author{T.~Huege$^{a}$,}
\author{P.G.~Isar$^{a}$,} 
\author{K.-H.~Kampert$^{d}$,} 
\author{D.~Kang$^{b}$,}
\author{D.~Kickelbick$^{i}$,}
\author{Y.~Kolotaev$^{i}$,}
\author{O.~Kr\"omer$^{c}$,} 
\author{J.~Kuijpers$^{e}$,} 
\author{S.~Lafebre$^{e}$,} 
\author{P.~{\L}uczak$^{l}$,}
\author{H.J.~Mathes$^{a}$,} 
\author{H.J.~Mayer$^{a}$,} 
\author{J.~Milke$^{a}$,}
\author{B.~Mitrica$^{h}$,} 
\author{C.~Morello$^{j}$,} 
\author{G.~Navarra$^{f}$,} 
\author{S.~Nehls$^{a}$,}
\author{A.~Nigl$^{e}$,} 
\author{J.~Oehlschl\"ager$^{a}$,} 
\author{S.~Over$^{i}$,}
\author{M.~Petcu$^{h}$,}
\author{T.~Pierog$^{a}$,} 
\author{J.~Rautenberg$^{d}$,} 
\author{H.~Rebel$^{a}$,} 
\author{M.~Roth$^{a}$,} 
\author{A.~Saftoiu$^{h}$,}
\author{H.~Schieler$^{a}$,}
\author{A.~Schmidt$^{c}$,}
\author{F.~Schr\"oder$^{a}$,}
\author{O.~Sima$^{m}$,} 
\author{K.~Singh$^{e,3}$,} 
\author{M.~St\"umpert$^{b}$,} 
\author{G.~Toma$^{h}$,}
\author{G.C.~Trinchero$^{j}$,} 
\author{H.~Ulrich$^{a}$,} 
\author{W.~Walkowiak$^{i}$,} 
\author{A.~Weindl$^{a}$,} 
\author{J.~Wochele$^{a}$,} 
\author{M.~Wommer$^{a}$,} 
\author{J.~Zabierowski$^{l}$,}
\author{J.A.~Zensus$^{g}$}

\vspace{2mm}
\address{
$^{a}$ Institut\ f\"ur Kernphysik, Forschungszentrum Karlsruhe, Germany\\
$^{b}$ Institut f\"ur Experimentelle Kernphysik, Universit\"at Karlsruhe, Germany\\
$^{c}$ Inst. Prozessdatenverarbeitung und Elektronik, Forschungszentrum Karlsruhe, Germany\\
$^{d}$ Fachbereich Physik, Universit\"at Wuppertal, Germany\\
$^{e}$ Dept. of Astrophysics, Radboud University Nijmegen, The Netherlands\\
$^{f}$ Dipartimento di Fisica Generale dell'Universit{\`a} Torino, Italy\\
$^{g}$ Max-Planck-Institut f\"ur Radioastronomie Bonn, Germany\\
$^{h}$ National Institute of Physics and Nuclear Engineering Bucharest, Romania\\
$^{i}$ Fachbereich Physik, Universit\"at Siegen, Germany\\
$^{j}$ Istituto di Fisica dello Spazio Interplanetario, INAF Torino, Italy\\
$^{k}$ ASTRON, Dwingeloo, The Netherlands\\
$^{l}$ Soltan Institute for Nuclear Studies Lodz, Poland\\
$^{m}$ Department of Physics, University of Bucharest, Romania\\
\vspace{0.5mm}
$^{\star}$ corresponding author: andreas.haungs@ik.fzk.de \\
$^{1}$ now at: Universidad Michoacana, Morelia, Mexico \\
$^{2}$ now at: Universidade S$\tilde{a}$o Paulo, Instituto de Fisica de S$\tilde{a}$o Carlos, Brasil\\
$^{3}$ now at: KVI, University of Groningen, The Netherlands  
}


\begin{abstract}
LOPES is set up at the location of the KASCADE-Grande extensive air shower
experiment in Karlsruhe, Germany and aims to measure and investigate radio pulses from 
Extensive Air Showers. Since radio waves suffer very little attenuation, 
radio measurements allow the detection of very distant or highly inclined
showers. 
These waves can be recorded day and night, and provide a bolometric measure of
the leptonic shower component. 
LOPES is designed as a digital radio interferometer using high bandwidths and 
fast data processing and profits from the reconstructed air shower observables 
of KASCADE-Grande. 
The LOPES antennas are absolutely amplitude calibrated allowing to reconstruct 
the electric field 
strength which can be compared with predictions from detailed Monte Carlo 
simulations. 
We report about the analysis of correlations present in the radio signals 
measured by the LOPES 30 antenna array. 
Additionally, LOPES operates antennas of a different type (LOPES$^{\rm STAR}$)
which are optimized for an 
application at the Pierre Auger Observatory. Status, recent results of the 
data analysis and further 
perspectives of LOPES and the possible large scale application of this new 
detection technique are discussed.  
\end{abstract}

\end{frontmatter}

\section{Introduction}

The traditional method to study extensive air showers (EAS), which 
are generated by high-energy cosmic rays entering the Earth's atmosphere, 
is to measure the secondary particles with sufficiently large particle 
detector arrays. In general these measurements provide only 
immediate information on the status of the air shower cascade 
on the particular observation level. This hampers the determination 
of the properties of the EAS inducing primary as compared to 
methods like the observation of Cherenkov and fluorescence light, 
which provide also some information on the longitudinal EAS 
development, thus enabling a more reliable access to the intended  
information~\cite{rpp}. 

In order to reduce the statistical and systematic uncertainties of 
the detection and reconstruction of EAS, especially with respect 
to the detection of cosmic particles of highest energies, there is 
a current methodical discussion on new detection techniques. 
Due to technical restrictions in past times 
the radio emission accompanying cosmic ray air 
showers was a somewhat neglected EAS feature. For a review on the
early investigations of the radio emission in EAS in the 
60ties see~\cite{Allan71}. 
However, the study of this EAS component has experienced a 
revival by recent activities, in particular by the LOPES project, and
the CODALEMA experiment in France~\cite{codalema}. 

The main goal of the investigations in Karlsruhe in the frame of  
LOPES is the understanding of the shower radio emission 
in the primary energy range of $10^{16}\,$eV to $10^{18}\,$eV. 
I.e., to investigate in detail the correlation of the measured 
field strength with the shower parameters, in particular the 
orientation of the shower axis (geomagnetic angle, azimuth angle, 
zenith angle), the position of the observer (lateral extension 
and polarization of the radio signal), and the energy and mass 
(electron and muon number) of the primary particle. Another goal 
of LOPES is the optimization of the hardware (antenna design and 
electronics) for a large scale application of the detection 
technique including a self-trigger mechanism for a stand-alone 
radio operation~\cite{gemmeke08}. 

Finally, within the frame of LOPES a detailed 
Monte-Carlo simulation program package is developed. 
The emission mechanism utilized in the REAS code 
(see~\cite{huege07} and references therein) is embedded in the 
scheme of coherent geosynchrotron radiation.
Progress in theory and simulation of the radio emission in air showers  
is described in a further contribution to this
conference~\cite{huege08}.

The present contribution sketches briefly recent results of the LOPES 
project~\cite{Falcke05} obtained by analyzing the correlations 
of radio data with shower parameters reconstructed by 
KASCADE-Grande~\cite{navarra,kascade}. 
Hence, LOPES, which is designed as digital radio interferometer using 
large bandwidths and fast data processing, profits from the 
reconstructed air shower observables of KASCADE-Grande.

\section{General layout, calibration, and data processing}

\subsection{General layout}

The basic idea of the LOPES (= LOFAR prototype station) project 
was to build an array of relatively simple, quasi-omnidirectional 
dipole antennas, where the received waves are digitized and sent 
to a central computer. This combines the advantages of low-gain 
antennas, such as the large field of view, with those of 
high-gain antennas, like the high sensitivity and good background 
suppression. 
With LOPES it is possible to store the received data stream for a 
certain period of time, i.e.~after a detection of a transient 
phenomenon like an air shower a beam in the desired direction 
can be formed in retrospect.
To demonstrate the capability to measure air showers with 
such antennas, LOPES is built-up 
at the air shower experiment KASCADE-Grande~\cite{navarra}. 
KASCADE-Grande is an extension of the multi-detector setup 
KASCADE~\cite{kascade}
(KArlsruhe Shower Core and Array DEtector) built in Germany, 
measuring the charged particles of air showers in the primary 
energy range of $100\,$TeV to $1\,$EeV with high precision due to 
the detection of the electromagnetic and the muonic shower component 
separately with independent detector systems. 
Hence, on the one hand LOPES profits from the reconstructed air 
shower observables of KASCADE-Grande, but on the other hand since 
radio emission arises from different
phases of the EAS development, LOPES also provides complementary 
information and helps to understand the observables measured with 
the particle detector arrays of KASCADE-Grande.

In the current (2007 and 2008) status LOPES~\cite{Horn06} 
operate 30 short dipole radio 
antennas (LOPES~30) and 10 logarithmic periodic dipole antennas
(LOPES$^{\rm STAR}$). The latter operate in both polarization
directions each (i.e.~20 channels), and are used for the 
development of a radio self-trigger system (see ref.~\cite{gemmeke08}). 

\begin{figure}
\begin{center}
\includegraphics*[width=6.2cm]{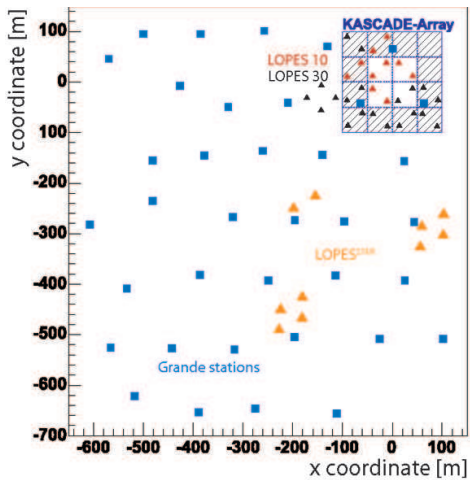}
\end{center}
\caption{Sketch of the KASCADE-Grande -- LOPES 
experiments: The 16 clusters (12 with muon counters) of the 
KASCADE field array, the distribution of the 37 stations of the Grande
array are shown. The location of the 30 LOPES radio antennas is also 
displayed as well as the positions of the 10 newly 
developed LOPES$^{\rm STAR}$ antennas.}
\label{FigLay}
\end{figure}

The LOPES~30 antennas, positioned within or close to the original 
KASCADE array (fig.~\ref{FigLay}), operate in the frequency range of 
$40-80\,$MHz and are aligned in east-west direction, i.e.
they are sensitive to the linear 
east-west polarized component of the radiation only, 
which can be easily changed into the opposite polarization 
by turning the antennas.
The layout provides the possibility for, e.g. a
detailed investigation of the lateral extension of the radio 
signal as LOPES~30 has a maximum baseline of approximately~$260\,$m.
The read out window for each antenna is $0.8\,$ms wide, 
centered around the trigger received from the KASCADE array. 
The sampling rate is $80\,$MHz.
The shape of the antenna and their metal ground screen gives the 
highest sensitivity to the zenith and half sensitivity to a zenith 
angle of $45^\circ$, almost independent on the azimuth angle.  
The logical condition for the LOPES-trigger is a high multiplicity of
fired stations of the KASCADE-Grande arrays. 
This corresponds to primary energies above $\approx 10^{16}\,$eV; 
such showers are detected at a rate of $\approx 2$ per minute. 

\subsection{Amplitude calibration}

\begin{figure}
\begin{center}
\includegraphics*[width=7cm]{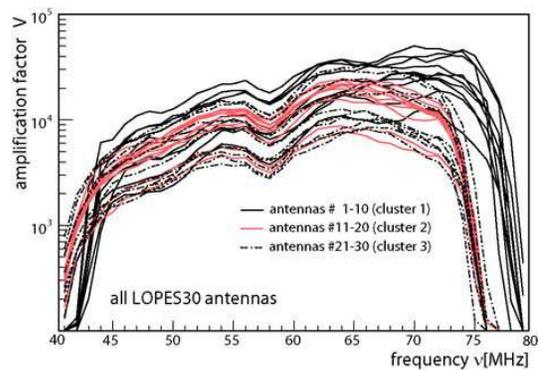}
\end{center}
\caption{Frequency dependent amplification factors for all 30 
antennas obtained by the amplitude calibration~\cite{Nehls08}.}
\label{FigCal}
\end{figure}

Each single LOPES~30 radio antenna is absolute amplitude calibrated at its 
location inside the KASCADE-Array (end-to-end calibration) using a 
commercial reference antenna~\cite{Nehls08} of known electric 
field strength at a certain distance. 
The power to be received from the source in 
calibration mode is compared with the power recorded in
the LOPES electronics. 
The calibration procedure leads to frequency dependent amplification
factors representing the complete system behavior (antenna, cables, electronics)
in the environment of the KASCADE-Grande experiment (fig.~\ref{FigCal}). 
The calibration results in a spread for the amplification factors 
between different antennas of nearly one order of magnitude.
The obtained correction factors are applied to the 
measured signal strengths resulting in the true electric field strength
which can be compared to simulated values.

\begin{figure}
\begin{center}
\includegraphics*[width=6.0cm]{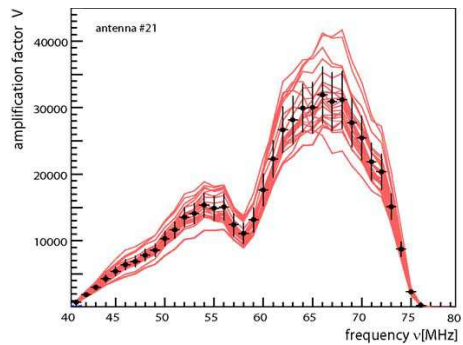}
\end{center}
\caption{Amplification factors (in linear scale) of one LOPES antenna for 
independent measurements. These measurements were spread over the course 
of nearly two years~\cite{Nehls08}.}
\label{FigCalS}
\end{figure}

Detailed investigations of possible sources of systematic uncertainties 
lead to a total uncertainty of $\sigma_V/V=70$~\% for the power 
(i.e.~$\sqrt{70}$\% for the electric field) related 
amplification factor averaged over the effective frequency range. 
The main contribution is due to a systematic uncertainty in the field strength of 
the used reference source itself (${\rm sys_{reference}}=67$~\%) as reported by the 
data sheet of the commercial antenna. Using another, more precisely 
calibrated reference radio source would improve this accuracy.
The uncertainty of the calibration method itself (${\rm sys_{calib}}=20.5$~\%) 
can be estimated from repeated measurement campaigns for a single 
antenna under all kinds of conditions. This total uncertainty also
includes e.g. environmental effects, like those caused by different
weather conditions present over the two years of calibration
campaigns (${\rm sys_{env}}=13$~\%) (fig.~\ref{FigCalS}).
The systematic uncertainty ${\rm sys_{calib}}$ of the calibration procedure 
has two additional sources: 
Electronic modules are temperature dependent and we have shown that
there is a relation between air temperature and amplification factor
$V(\nu)$ for the LOPES antenna system. A more precise correlation
analysis and following correction can improve the overall precision
for measuring electric field strengths.
The antenna gain simulation contributes with large amount (15\%) to the total
uncertainty. In particular the simulation reveals a resonance at 58~MHz,
where the measurements show that it is less pronounced and should be 
re-evaluated or interpolated in the gain calculations~\cite{Nehls08}. 

In addition, during LOPES measurements we place emphasis on the monitoring
of the environmental conditions by measuring the static electric 
field and by recording parameters of nearby weather stations. 
Atmospheric conditions, in particular E-field
variations during thunderstorms influence the radio emission 
during the shower development and the measurement of the radio pulses. 
By monitoring the environmental conditions,
and comparing them with the antenna noise level as well as with the detected air
shower radio signals, correlations will be investigated and corrected for.

An important conclusion is that the discussed strategy of calibration
can be adapted for future radio antenna arrays measuring cosmic ray
air showers. Especially at locations with much lower RFI an
astronomical source, e.g. the galactic background radiation, can be
used to cross-check the proposed amplitude calibration of a radio
antenna system.

\subsection{Data processing}

The LOPES data processing includes several steps. 
First, the relative instrumental delays are corrected using a 
known TV transmitter visible in the data. 
End of 2007 this TV station was switched off, but could be 
replaced by a local dedicated source on site the Karlsruhe 
research center. The TV- as well as now the local source are also used 
for time calibration issues of LOPES. 
Next, the digital filtering, gain corrections and corrections 
of the trigger delays based on the known shower direction 
(from KASCADE-Grande) are applied and noisy antennas are flagged. 

The geometrical delay (in addition to the instrumental delay 
corrections) by which the data is shifted, is the time difference 
of the pulse coming from the given direction to reach 
the position of the corresponding antenna compared to the reference 
position. This shift is done by multiplying a phase gradient 
in the frequency domain before transforming the data back to 
the time domain.
This step includes the application of the calibration 
correction factors and also a correction for the azimuth and zenith
dependence of the antenna gain. 

To form the beam from the time shifted data, the data from each 
pair of antennas is multiplied time-bin by time-bin, the resulting 
values are averaged, and then the square root is taken while 
preserving the sign.
We call this the cross-correlation beam or CC-beam.

The quantification of the radio parameters is by fitting the 
CC-beam pulse:
Although the shape of the resulting pulse (CC-beam) is not really 
Gaussian, fitting a Gaussian to the smoothed data gives
a robust value for the peak strength, which is defined as the
height of this Gaussian. 
The error of the fit results gives also a first estimate of the 
uncertainty of this parameter. 
The finally obtained value $\epsilon_{\rm CC}$, which is the 
measured amplitude divided by the effective bandwidth, is
used for the analysis. 

The analysis of the data using the CC-beam is based on the RFI cleaned
raw data. However, the sampling of the data is done in the second
Nyquist domain and a reconstruction of the original 40--80~MHz signal
shape is needed to investigate the radio emission properties in more details.
Therefore, an up-sampling of the data on a single antenna basis is performed 
(by the zero-padding method applied in the frequency domain) resulting in a 
band limited interpolation in the time domain~\cite{Nehls08a} 
to reconstruct the original signal form between the sampled data 
points with 12.5~ns spacing. 
The method can be applied, because the needed information
after sampling in the second Nyquist domain are contained in the
stored data~\cite{Kroem08}.

After applying the up-sampling and subtraction of an estimated noise level, 
the radio signals can be used to reconstruct the electric field strength 
in each individual antenna.
By technical reasons, the CC-beam calculation could not been used on the 
up-sampled data at the time of this analysis. Therefore, for the 
following results up-sampling is only 
performed when signals in single antennas are discussed.

\section{Results from the initial 10 LOPES antennas}

First measurements (6 months data taking) were performed with a set 
up of 10 antennas (LOPES~10) with remarkable results~\cite{Haungs06}.
With a sample asking for high quality events 
the proof-of-principle for detection of air showers in the radio 
frequency range was made~\cite{Falcke05}.
In total, more than 600 events with a clear radio signal and with 
the shower core inside the KASCADE-Grande experiment could be detected. 
The analysis of these events concentrated on the correlations 
of the radio signal with shower parameters, 
in particular with the arrival direction and with the shower size, 
i.e. the primary energy of the shower.    

The results support the expectation that the field strength 
increases by a power-law with an index close to one
with the primary energy, i.e. that the received energy of the 
radio signal increases quadratically with the primary energy of the 
cosmic rays~\cite{Horn06}. An index of this power-law close to one 
serve as a proof of the coherence of the radio emission 
during the shower development. 
A clear correlation was also found with the geomagnetic angle 
(angle between shower axis and geomagnetic field direction) which 
indicates a geomagnetic origin for the emission mechanisms.    

Due to the geometrical layout of the KASCADE-Grande array 
(see fig.~\ref{FigLay}) the radio signal can be investigated for events 
which have distances up to $800\,$m from the center of the antenna setup.
Investigating the average lateral behavior of 
the radio emission in more detail, a clear correlation of the signal 
strength with the mean distance of the shower axis to the antennas was found. 
By assuming an exponential function, the scaling parameter 
$R_0$ resulted to $230\pm51\,$m~\cite{Badea06}.

Further interesting features of the radio emission in EAS were 
investigated with a sample of very inclined showers~\cite{Petrovic07} 
and with a sample of events measured during thunderstorms~\cite{Buitink07}. 
The first one is of special interest for a large scale application 
of this detection technique, as due to the low attenuation in the 
atmosphere also very inclined showers can be detected with high efficiency. 
This is of great importance if ultrahigh energy neutrinos exist. 
With LOPES one could show that events above $70^\circ$ zenith angle
still emit a detectable radio signal. An update of this analysis is presented
in ref.~\cite{Saftoiu08}.
The latter sample is of interest to investigate the role of the 
atmospheric electric field in the emission process. 
LOPES data were recorded during thunderstorms, periods of heavy cloudiness 
and periods of cloudless weather. It was found that during thunderstorms the radio 
emission can be strongly enhanced, where no amplified pulses were found 
during periods of cloudless sky or heavy cloudiness, suggesting that the 
electric field effect for radio air shower measurements can be safely ignored
during non-thunderstorm conditions~\cite{Buitink07}.

\section{Results of LOPES~30}

\subsection{Correlation with primary energy}

\begin{figure}
\begin{center}
\includegraphics*[width=7cm]{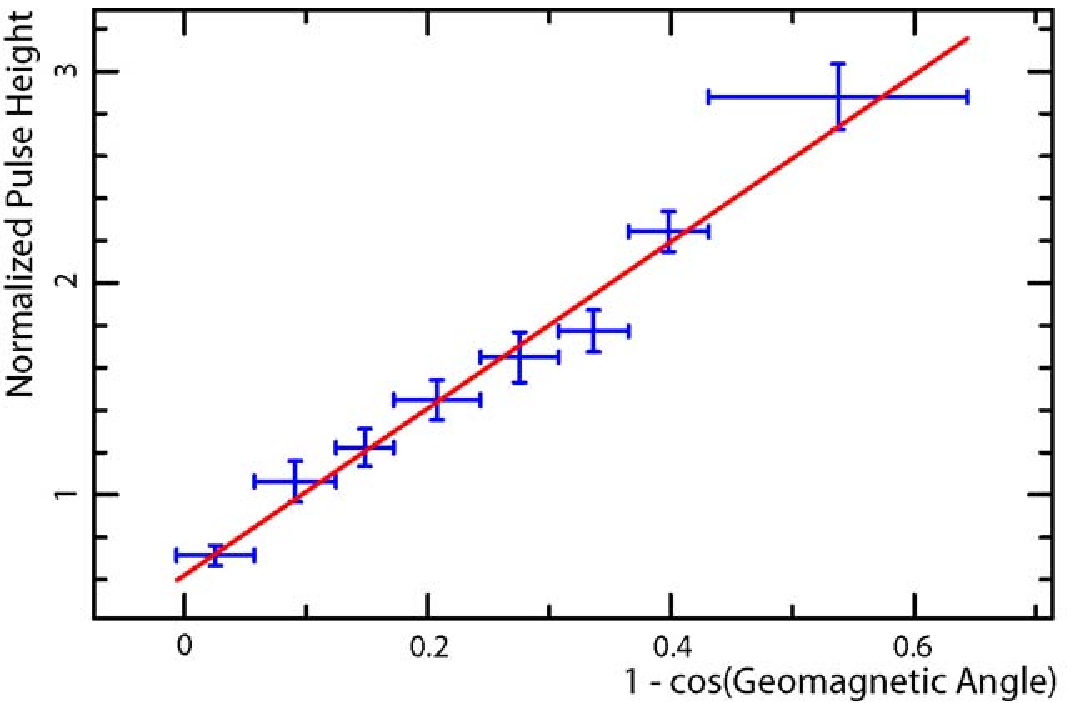}
\includegraphics*[width=7cm]{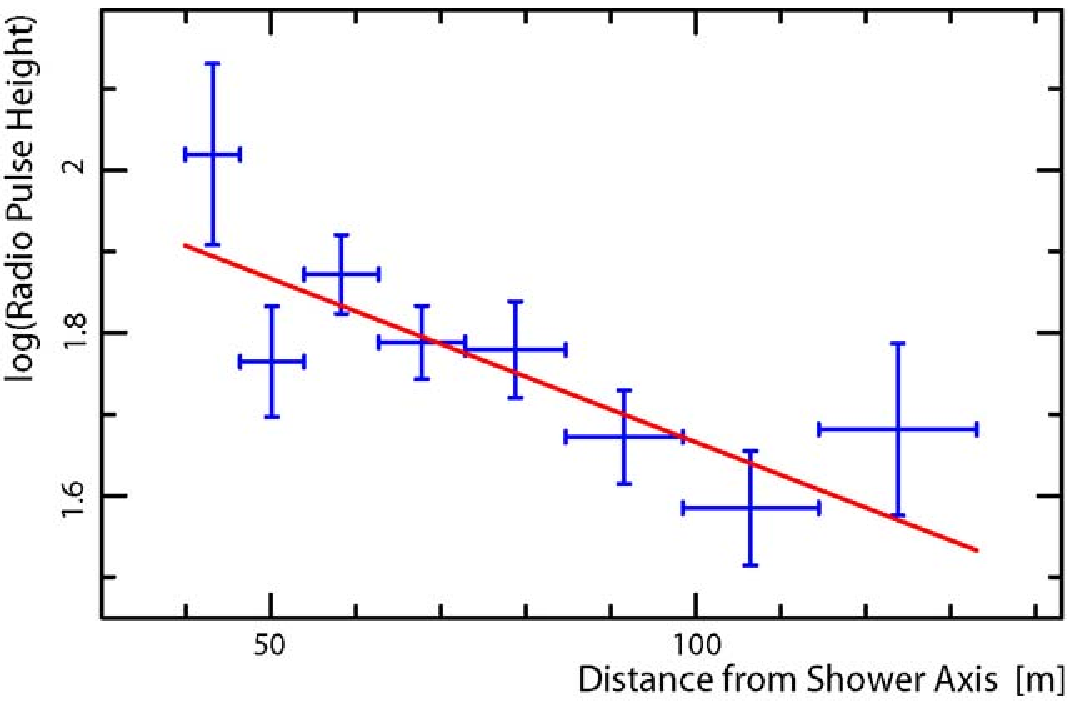}
\includegraphics*[width=7cm]{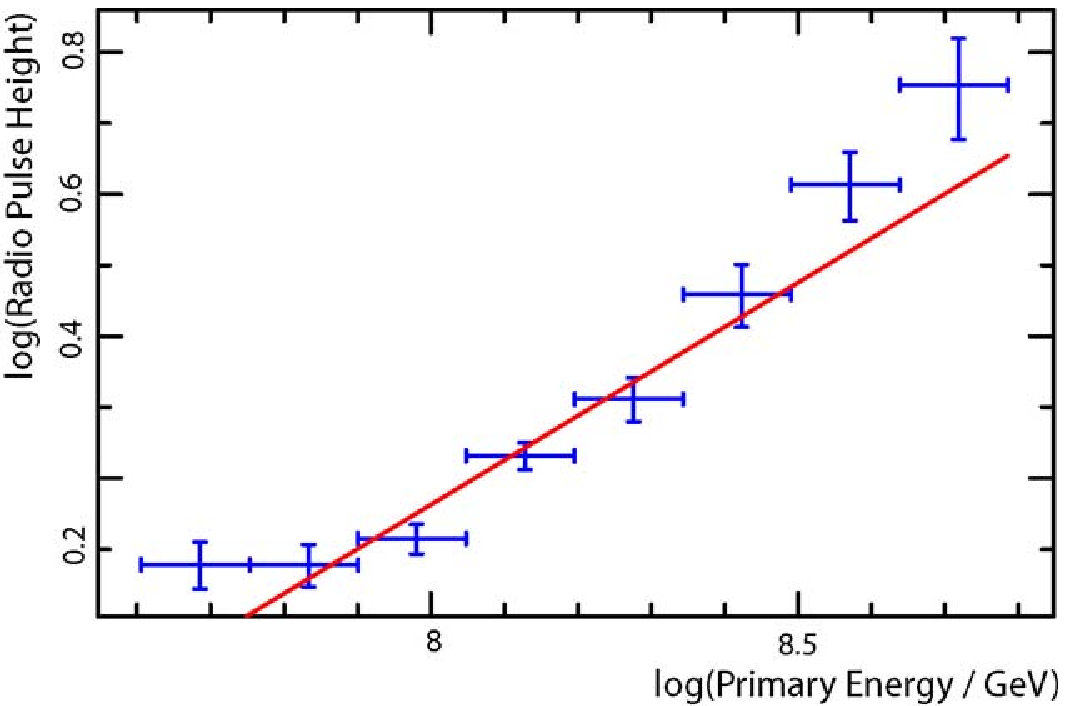}
\end{center}
\caption{Normalized radio pulse height plotted versus 
(from top to bottom):
the angle to the geomagnetic field, 
the mean distance of the antennas to the shower axis,
and the estimated primary particle energy~\cite{Horn07}.
The normalization is done by dividing by the fits to the other parameters 
(geomagnetic angle, distance, or energy). }
\label{Figpara}
\end{figure}

The radio pulse height (CC-beam) measured by the 30 east-west polarized antennas 
of LOPES~30 can be parameterized as a function of the angle to the geomagnetic field, 
the zenith angle, the distance of the antennas to the air shower axis and 
an estimate of the primary particle energy calculated from 
KASCADE-Grande data~\cite{Horn07}. 
The separated relations for the LOPES~30 events are displayed in 
figure~\ref{Figpara}.
The fits to the geomagnetic angle, the distance to the shower axis, and to the primary 
energy are done separately and results in an analytical expression for the radio pulse height based on the estimated shower observables:
\begin{eqnarray}
\nonumber
&\epsilon_{\rm est}  =
(11\pm1.)
\left((1.16\pm0.025)-\cos\alpha\right) \cos\theta & \nonumber \\
& \exp\left(\frac{\rm -R_{SA}}{\rm (236\pm81)\,m}\right)
\left(\frac{\rm E_{p}}{\rm 10^{17}eV}\right)^{(0.95\pm0.04)} 
\left[\frac{\rm \mu V}{\rm m\,MHz}\right]&
\label{eq:horneffer-energy}
\end{eqnarray}
(With: $\alpha$ the geomagnetic angle, $\theta$ the zenith angle,
${\rm R_{SA}}$ the mean distance of the antennas to the shower axis,
and ${\rm E_{p}}$ the primary particle energy. The given errors are the 
statistical errors from the fit.) 
One issue that has to be kept in mind is that this analysis is only made with
the east-west polarized component, which can be the reason for the
debatable $1-\cos\alpha$ dependence on the geomagnetic angle.

The analytical formula derived from LOPES measurments can be inverted and allows then 
an estimate of the primary particle energy from radio data. 
The combined statistical spread for the estimation of the energy of single 
events from LOPES data and KASCADE-Grande data is 27\% for strong events.
This is in the same range as the fluctuations in measurements with particle 
detector arrays alone.

\subsection{Direction resolution}

\begin{figure}
\begin{center}
\includegraphics*[width=7cm]{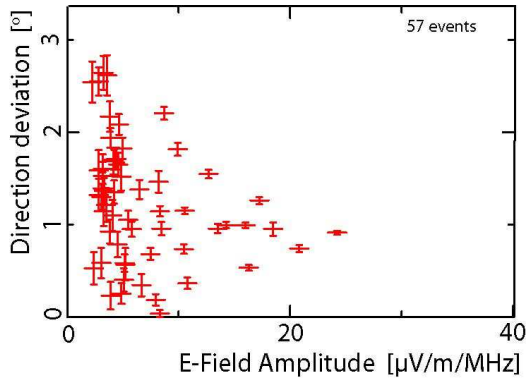}
\end{center}
\caption{Absolute angular separation between the
LOPES and the KASCADE position as a function of electric
field amplitude~\cite{Nigl08}. The statistical uncertainty in the
direction of each event is plotted as an error bar.}
\label{FigDir}
\end{figure}

To investigate the capabilities of measuring radio signals in terms of direction estimates, 
we produce 4-dimensional radio images on time-scales of nanoseconds using the digital 
beam-forming~\cite{Nigl08}. 
We search this multi-dimensional parameter space for the direction of maximum 
coherence of the air shower radio signal and compare it to the direction provided by
KASCADE.
Each pixel of the image is calculated for three
spatial dimensions and as a function of time. The third spatial dimension is obtained by 
calculating the beam focus for a range of curvature radii
fitted to the signal wave front. 
We find that the direction of the emission maximum can change when optimizing 
the curvature radius. This dependence dominates the statistical
uncertainty for the direction determination with LOPES. 
Furthermore, we find a tentative increase of the curvature radius to
lower elevations, where the air showers pass through a larger atmospheric depth. 
The distribution of the offsets between the directions of both installations is found to 
decrease linearly with increasing signal-to-noise ratio (fig.~\ref{FigDir}). 
We conclude that the angular resolution of LOPES is sufficient to determine the direction 
which maximizes the observed electric field amplitude. 
However, the statistical uncertainty of the directions is not determined by the resolution
of LOPES, but by the uncertainty of assuming a pure spheric wave. 
In addition, there are no systematic deviations between the directions
determined from the radio signal and from the detected particles. 

\subsection{Frequency spectrum}
\begin{figure}
\begin{center}
\includegraphics*[width=7cm]{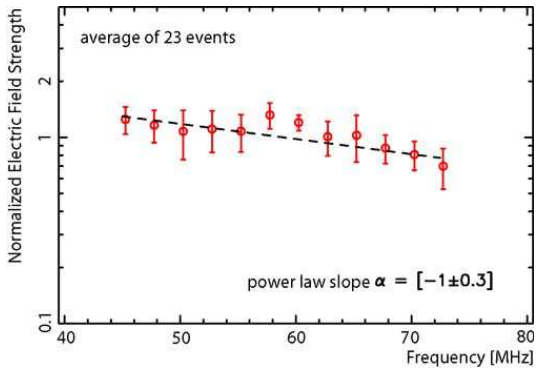}
\end{center}
\caption{Comparison of average cosmic-ray electric
field spectra obtained with 23 LOPES events. 
The frequency bin values determined on the CC-beam are 
fitted with a power law function (dashed line)~\cite{Nigl08a}.}
\label{FigSpec}
\end{figure}
The quality of the LOPES data allows us to study the spectral dependence of 
the radio emission from cosmic-ray air showers around 100 PeV.
With a sample of 23 strong LOPES events, the radio spectrum received 
from cosmic-ray air showers in the east-west polarization direction over 
a frequency band of 40 MHz is analyzed.
The radio data are digitally beam-formed before the spectra are determined 
by sub-band filtering~\cite{Nigl08a}.

The resulting electric field spectra fall off to higher frequencies 
for all individual events. 
The spectral slopes of the selected sample
of events depend on the length of the pulse, where
longer pulses result in steeper spectra. However, the spectra
do not show a significant dependence of the slope on
the electric field amplitude, the azimuth angle, the zenith
angle, the curvature radius, nor the average distance of
the antennas to the shower core positions.
The accuracy, with which the spectral amplitudes
can be obtained, is limited by the instrument
noise and phase uncertainties. Furthermore, the quality
of the spectral slope is limited by the quality of the antenna
gain model, which was simulated and measured in
several calibration campaigns~\cite{Nehls08}.

The average electric field spectrum of the 23 events can be fitted with 
an exponential function $\epsilon_\nu = K \cdot \exp{(\nu/{\rm MHz}/\beta)}$ and 
$\beta = -0.017 \pm 0.004$, or alternatively, with a power law 
$\epsilon_\nu = K \cdot \nu^\alpha$ and a spectral index
of $\alpha = -1.0 \pm 0.3$ (fig.~\ref{FigSpec}). 
The average slope of the spectra obtained with LOPES
confirms basic expectations, but it is slightly steeper than
the slope obtained from Monte Carlo simulations based on
coherent geosynchrotron emission from fully parametrized
air showers~\cite{Nigl08a}. 

\subsection{Lateral extension}

\begin{figure}
\begin{center}
\includegraphics*[width=7cm]{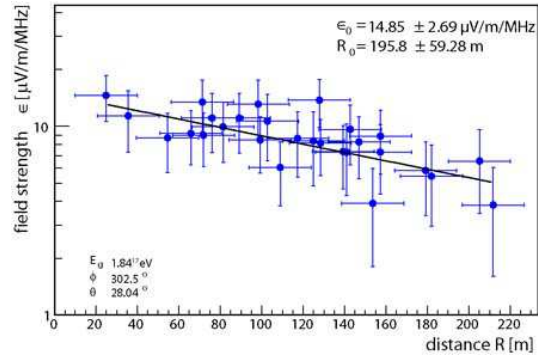}
\end{center}
\caption{Lateral distribution reconstructed from single antenna signal, 
shown for an individual shower~\cite{Nehls08a}.}
\label{FigLatex}
\end{figure}

For the analysis of lateral distributions of the radio emission in individual 
events 110 showers with a high signal-to-noise ratio were selected. 
Up-sampling is used to derive the electric field strength $\epsilon$ for
individual antennas. 
A background subtraction is performed based on a calculation using 
a time window (520 nanosecond width) before the actual radio pulse from
the shower.  
The distance of the antennas to the shower axis is obtained with help of 
the reconstructed shower parameters from KASCADE-Grande. 
To investigate the lateral behavior of the radio signal an exponential function 
$\epsilon=\epsilon_0\cdot\exp\left( -R/R_0 \right)$ was used to describe 
the measured field strengths. The fit contains two free parameters, where the scale 
parameter $R_0$ describes the lateral profile and $\epsilon_0$ the extrapolated 
field strength at the shower axis at observation level.
An example of an individually measured event including the resulting lateral field strength 
function is shown in figure~\ref{FigLatex}.

\begin{figure}
\begin{center}
\includegraphics*[width=7cm]{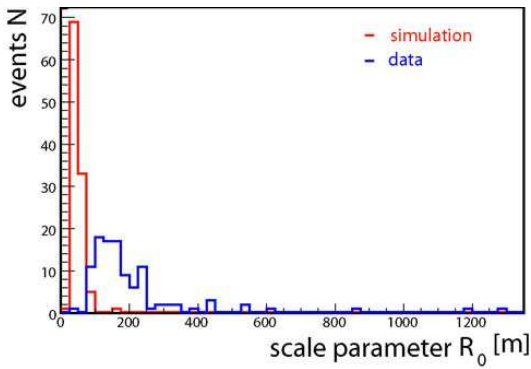}
\end{center}
\caption{Distribution of the scale parameter $R_0$ obtained from
measurements and simulations~\cite{Nehls08a}.}
\label{FigLatdis}
\end{figure}

From the distribution of the obtained scale parameters
(figure~\ref{FigLatdis}) one derives that the scale parameter
peaks at $R_0\approx125$~m and has a mean value of $R_0=237$~m for the
detected events. Here, the distribution comprises showers with an
expected exponential decay as well as events with a very flat
lateral distribution. Roughly 10\% of the investigated showers 
show very flat lateral distributions with very large scale parameters.

The found lateral distributions with very flat lateral profiles
are remarkable and require further investigations with higher statistics. 
The fact that we measured lateral distributions with a flat behavior towards 
the shower center or even over the whole observable distance range, can not be
simply explained with instrumental effects.

Including the tail of the
distribution in figure~\ref{FigLatdis} the mean value agrees
with the ´CC-beam based scale parameter in the parameterization of
\cite{Horn07}, whereas an exclusion of the tail obtains a scale
parameter that agrees with the parameterization of~\cite{Allan71}.
\begin{figure}
\begin{center}
\includegraphics*[width=7cm]{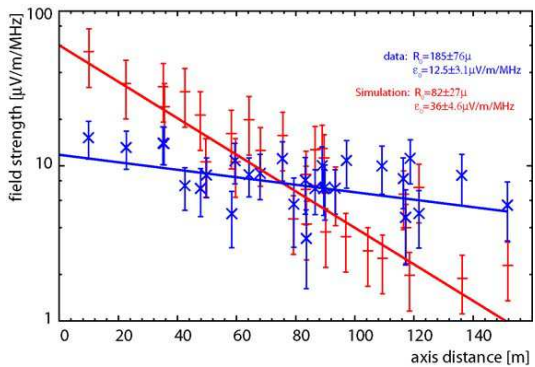}
\end{center}
\caption{Lateral distribution obtained for data
(blue) and simulation (red) for an individual event~\cite{Nehls08a}.
}
\label{FigSimMeas}
\end{figure}

A very important comparison for the understanding of the radio emission
observed in EAS, the geosynchrotron radiation, was performed with 
detailed Monte Carlo simulations on an event-to-event basis. 
This is now possible, due to the performed amplitude calibration of 
LOPES and the estimate of the field strength at individual antennas. 
It is as a basic necessity
in order to get the absolute scale for the field strength. Due to a
new simulation strategy, using more realistic air shower models with
precise, multi-dimensional histograms derived from per-shower 
CORSIKA~\cite{cors} simulations, detailed comparisons can be performed.

The REAS2 Monte Carlo simulation code 
(see~\cite{huege07,huege08} and references therein) is used to
simulate the geosynchrotron radio emission for all the showers
detected in the investigated data set. For each single event 250
showers were simulated with the fast one-dimensional shower simulation program 
CONEX, using the same values for the direction 
and the guessed primary energy under the assumption of a proton induced 
shower. 
The shower that represents the mean of all 250 simulated showers
is selected. From the selected CONEX~\cite{conex} shower the particle 
stack after the first
interaction is used as input for a full CORSIKA simulation. 
The resulting information and the known shower core position 
is used in the REAS2 code to calculate 
the radio emission. The output are unlimited
bandwidth pulses, that are digitally filtered with a rectangle filter
from 43 to 76~MHz for the known antenna positions at ground, which can 
be directly compared with the measured lateral distribution 
(fig.~\ref{FigSimMeas}).  

The comparison of the distributions of the scale parameter from
measurement and simulation is shown in figure~\ref{FigLatdis}.
The mean for the distribution of the scale parameter from the simulation 
is $R_0=50$~m.
Such small values represent a steep lateral decrease of the field
strength. In general the simulations give steeper lateral distributions. 
In addition, it
was derived that the differences between measurements and simulations
can be very large and that the unexpected very flat lateral profiles
are never reproduced in simulations. In order to understand the
underlying systematic effects significantly more statistics is needed.

\begin{figure}
\begin{center}
\includegraphics*[width=7cm]{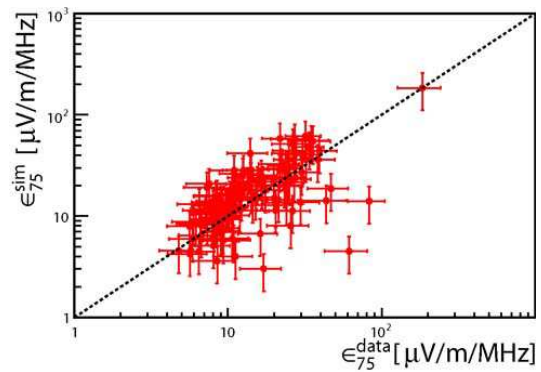}
\end{center}
\caption{Comparison of field strength at distance $R$.
    Correlations $\epsilon_{R}^{\rm data}$ obtained from
    measurements and $\epsilon_{R}^{\rm sim}$ obtained from
    simulation. The dashed line represents equal values from
    simulation and measurement~\cite{Nehls08a}.
 }
\label{FigEpsilon}
\end{figure}

The deviation in the scale parameters enters in 
systematically lower field strengths $\epsilon_0^{\rm data}$,
compared to the field strengths $\epsilon_0^{\rm sim}$ obtained from 
simulations at the shower center. The difference is a factor of three 
at the shower center. On the other hand we obtained at $R=75$~m a 
fairly good agreement between simulations and measurements 
(fig.~\ref{FigEpsilon}) for all events. This is a 
promising result in itself, as such comparisons are performed for the 
first time for LOPES data.

\subsection{Polarization measurements}
After one year of measurements of the east-west polarization component 
by all 30 antennas, the LOPES~30 set-up was reconfigured to perform dual-polarization
measurements. Half of the antennas have been configured for measurements of the 
north-south polarization direction. By measuring at the same time both polarization 
components and by comparing with the expectations, the geosynchrotron effect as the 
dominant emission mechanism in air showers will be experimentally verified.
First results on the analysis of these data are discussed in reference~\cite{Isar08}.

\section{Large scale application: LOPES$^{\rm STAR}$}

One of the main goals of the LOPES project is to pave the way
for an application of this `re-discovered' air shower detection
technique in large UHECR experiments, like the
Pierre Auger Observatory. 
Parallel to the measurements at KASCADE-Grande LOPES follows 
this aim by optimizing the antenna design for an application 
at Auger, named LOPES$^{\rm STAR}$~\cite{gemmeke08}.
Going in direction of setting up a test array at Auger South, the
possibilities of new antenna types and in particular, a self-triggering 
antenna system are also investigated.
Meanwhile several dual polarized STAR-antennas are in operation at 
the KASCADE-Grande field (see fig.~\ref{FigLay}). Also by this system air showers 
could be detected. This information is used to optimize the 
self-trigger system of LOPES$^{\rm STAR}$.
In parallel to the activities in Karlsruhe, also the first test set up 
with LOPES$^{\rm STAR}$ antennas in Argentina at the Auger South experiment 
has detected first radio signals from extensive air-showers~\cite{Coppens08}.

\section{Summary}

The main goals of the LOPES project are the investigation of 
the relation between the radio emission from extensive 
air showers with the properties of the primary particles and the 
development of a robust, autonomous, and self-triggering antenna set-up 
usable for large scale applications of the radio detection technique 
of air-showers.

With LOPES~30 we are able to follow the first goal of
the LOPES project: The calibration of the radio emission 
in extensive air showers for primary energies below $10^{18}\,$eV.
Because of the absolute amplitude 
calibration a direct comparison of the field strength with the expectations 
(simulations) is possible.

With LOPES the proof-of-principle for detection of cosmic particles 
by radio flashes from extensive air showers could be performed. 
First results obtained by correlating the observed 
radio field strength with the shower parameters obtained 
by the KASCADE measurements appear to be very promising 
for a more detailed understanding of the emission mechanism 
from atmospheric showers.  
Most interesting results are the found quadratic dependence of the 
radio-power on energy, the correlation of the radio field strength 
with the direction of the geomagnetic field, the exponential behavior 
of the lateral decrease of the field strength with a scaling parameter
in the order of hundreds of meter, and that except during strong thunderstorms
the radio signal is not strongly influenced by weather conditions.  
Large scaling radii allow us to measure the same field strength at
larger distances from the shower core, which will be helpful for
large scale applications of the radio detection technique.
In particular, the quadratic dependence on energy will make radio detection 
a cost effective method for measuring the longitudinal 
development of air showers of the highest energy cosmic rays 
and cosmic neutrinos. This result places a strong supportive argument for the use
of the radio technique to study the origin of high-energy cosmic rays.

LOPES is still running and continuously takes data in coincidence with the air shower 
experiment KASCADE-Grande. In this frame also improvements in the technology 
and the development of the self-trigger concept are tested. 

Besides the experimental work done with the present antenna setup 
the LOPES project aims to improve the theoretical understanding of 
the radio emission in air showers. 
Supplementary emission processes like the 
Cherenkov-Askaryan-effect which plays the dominant 
role in dense media will be investigated.

The LOPES technology can be applied to existing cosmic ray
experiments as well as to large digital radio telescopes like LOFAR
and the SKA (square kilometer array), providing a large detection 
area for high energy cosmic rays. First approaches 
to use the technique at the Pierre Auger Observatory and 
at a first LOFAR station are under way.  

\ack{
LOPES was supported by the German Federal Ministry of Education
and Research. The KASCADE-Grande experiment is supported
by the German Federal Ministry of Education and
Research, the MIUR and INAF of Italy, the Polish Ministry
of Science and Higher Education and the Romanian Ministry
of Education and Research.}

\end{document}